# Optimizing the number of CNOT gates in one-dimensional nearest-neighbor quantum Fourier transform circuit

Byeongyong Park[1,2], Doyeol Ahn[1,2,3]*

[1]Department of Electrical and Computer Engineering and Center for Quantum Information Processing, University of Seoul, 163 Seoulsiripdae-ro, Dongdaemun-gu, Seoul 02504, Republic of Korea.

[2]First Quantum Inc., 2F-210, Sparkplus, 180, Bangbae-ro, Seocho-gu, Seoul 06586, Republic of Korea.

[3]Physics Department, Charles E. Schmidt College of Science, Florida Atlantic University, 777 Glades Road, Boca Raton, FL 33431-0991, USA.

*Corresponding author: dahn@uos.ac.kr

**Abstract:** The physical limitations of quantum hardware often require nearest-neighbor qubit structures, in which two-qubit gates are required to construct nearest-neighbor quantum circuits. However, two-qubit gates are considered a major cost of quantum circuits because of their high error rate as compared with single-qubit gates. The controlled-not (CNOT) gate is the typical choice of a two-qubit gate for universal quantum circuit implementation together with the set of single-qubit gates. In this study, we construct a one-dimensional nearest-neighbor circuit of quantum Fourier transform (QFT), which is one of the most frequently used quantum algorithms. Compared with previous studies on n-qubit one-dimensional nearest-neighbor QFT circuits, it is found that our method reduces the number of CNOT gates by $\sim 60\%$. Additionally, we showed that our results for the one-dimensional nearest-neighbor circuit can be applied to quantum amplitude estimation.

# I. INTRODUCTION

Quantum algorithms are of growing interest because the accelerated processing speed over classical algorithms for solving many practically important problems is expected [1–5]. On the other hand, quantum states, the subject of quantum algorithm executions, are very fragile because of the decoherence, which sometimes causes critical errors when implementing quantum algorithms. Most quantum algorithms are implemented using quantum circuits comprising qubits and gates. As decoherence results from the interaction with the external environment, the number of gate operations is the main cause of errors. Therefore, quantum circuits should be designed with minimal gates [6], especially in the noisy intermediate scale quantum (NISQ) era [7]. According to the Solovay-Kitaev theorem [8], universal quantum circuits are implemented by single-qubit and two-qubit gates such as controlled-not (CNOT) gates. The number of two-qubit gates is considered the main cost of quantum circuits in the area of NISQ computation because they cause more errors than single-qubit gates. Thus, optimizing the number of controlled-not (CNOT) gates would be a pivotal task.

For some near-term quantum computers like those of IBM or Rigetti, a fundamental requirement is that the physically implemented circuit must be the nearest-neighbor circuit since the connectivity between qubits is incomplete yet. Nearest-neighbor circuit means that a qubit in the circuit only interacts with adjacent qubits. For the general case, SWAP gates are used to make the circuit into a nearest-neighbor circuit. A SWAP gate is executed using 3 CNOT gates.

The quantum Fourier transform (QFT) is a crucial tool for many quantum algorithms such as quantum addition [9], quantum phase estimation [10], quantum amplitude estimation [3], the algorithm for solving linear systems of equations [4], and Shor's factoring algorithm [1].

The standard n-qubit QFT circuit requires $n(n-1)/2$ controlled-phase and n Hadamard (H) gates if reordering is allowed [11] [see Fig. 1(a)]. Reordering means that the order of qubits can be different before and after the circuit execution. To implement a controlled-phase gate, 2 CNOT and 3 $R_z$ gates are required [12]. Therefore, $n(n-1)$ CNOT gates are required to construct an n-qubit QFT circuit. If the nearest-neighbor quantum circuit is required to implement the QFT, the number of CNOT gates is much larger than $n(n-1)$. [13-17]

In this study, we investigate optimized n-qubit Quantum Fourier Transform (QFT) circuit design that has a one-dimensional nearest-neighbor structure by minimizing the number of CNOT gates. Previous studies [13-16] on one-dimensional nearest-neighbor QFT circuits focused on reordering the qubits of the standard QFT circuit using SWAP gates or the circuit of Ref. [17]. However, our circuit uses CNOT gates directly by decomposing the standard QFT circuit which would make our circuit more compact than that uses SWAP gates because implementing each SWAP gate needs three CNOT gates. Thus, our circuit design protocol would be a new starting point for optimizing the QFT circuit costs by replacing the standard QFT circuits or those in Ref. [17]. Upon reordering, our n-qubit one-dimensional nearest-neighbor QFT circuit requires $n^2 + n - 4$ CNOT gates, which are ~40% smaller than required in Ref. [17]. For 5- to 10-qubit QFT, our circuit has much lower CNOT counts than previous studies [15, 16] that used SWAP gates to make a one-dimensional nearest-neighbor QFT circuit. We also implement the quantum phase estimation (QPE) using the QFT circuit and showed how to apply our circuit construction method to the controlled-$R_y$ gates sharing the target qubit, which frequently appear in quantum amplitude estimation (QAE) which is expected to replace Monte Carlo integration [18, 19].

## II. MAIN RESULT

### A. Construction of nearest-neighbor quantum Fourier transform circuit

In this section, we construct a one-dimensional nearest-neighbor QFT circuit and reduce the number of required CNOT gates. The process is as follows: The standard QFT circuit in Fig. 1(a) can be decomposed into that in Fig. 1(b) using the methods proposed in Ref. [20]. By applying the circuit identity in Fig. 2 to the circuit in Fig. 1(b) and the cancelation between adjacent CNOT gates, we obtain the circuit in Fig. 3. Note that the circuit in Fig. 3 is a one-dimensional nearest-neighbor 5-qubit QFT circuit. Using the same method, we can construct an n-qubit one-dimensional nearest-neighbor QFT circuit that has $n^2 + n - 4$ CNOT gates which is much smaller than the previous results [15-17].

For the n-qubit QFT circuit, our circuit reduces the CNOT gates by ~60% when comparing the leading order terms with the previous study in Ref. [17]. For 5- to 10-qubit QFTs, our results reduce the number of CNOT gates by 16.13%, 20.83%, 30.67%, 43.80%, 47.88%, and 51.89% compared to the best-known results [15, 16] at the time of writing. Our result and comparison with previous works can be found in Table. I.

### B. Quantum phase estimation on the actual quantum hardware

In this section, we implemented QPE on actual quantum computers. QPE is an algorithm for finding an eigenvalue of a unitary operator using a corresponding eigenstate and QFT [11]. A detailed explanation of QPE can be found in Appendix A. Here, we chose the unitary operator U and the corresponding eigenvector $|u\rangle$ as below.

$$U = \begin{pmatrix} 1 & 0 \\ 0 & e^{2\pi i\theta} \end{pmatrix}, |u\rangle = \begin{pmatrix} 0 \\ 1 \end{pmatrix} \tag{1}$$

First, we implemented a 3-qubit QPE using Amazon Bracket and Rigetti-Aspen-11, which is a 40-qubit superconducting quantum computer. Next, we implemented a 3-qubit QPE using an IBM Quantum and ibm_nairobi, which is a 7-qubit superconducting quantum computer. The connectivities between qubits are shown in Fig. 4.

For both cases, we chose $\theta$ as 1/8, 2/8, 3/8, … 7/8, and implemented 3-qubit QPE using 4 qubits 1000 times for each $\theta$. Using Rigetti-Aspen-11, we obtained the right answer by taking a majority vote for all $\theta$. The probability to find the right answer was 26.23% on average. Furthermore, we also found the right answer using a majority vote for all $\theta$ using ibm_nairobi. The probability of finding the right answer was 47.6% on average. The result and comparison can be found in Fig. 5.

### C. Applying to quantum amplitude estimation

Our method to make a nearest-neighbor QFT circuit can be used for circuits of other quantum algorithms, such as QAE. QAE uses quantum amplitude amplification by the repetitive execution of a Grover operator. A circuit of controlled-$R_y$ gates sharing the target qubit frequently appears when QAE replaces the Monte Carlo integration [19, 21]. The explanation of QAE and the reason why controlled-$R_y$ gates are frequently used can be found in Appendix B.

To apply our method to the controlled-$R_y$ gates sharing the target qubit, the identity $R_x(-\pi/2)R_z(\theta)R_x(\pi/2) = R_y(\theta)$ is applied to the target qubit. Then, the controlled-$R_y$ gates are converted to controlled-$R_z$ gates, and our method to decompose QFT circuit is applied to make the circuit a nearest-neighbor circuit.

## III. CONCLUSION

In this study, we propose a novel one-dimensional nearest-neighbor n-qubit QFT circuit that reduces the number of CNOT gates by ~40% as compared with previous results [17]. Compared with previous studies [15, 16] on 5- to 10-qubit QFTs, our proposed circuit reduces the number of CNOT gates substantially when the circuit is a one-dimensional nearest-neighbor circuit. Moreover, we demonstrated that our circuit synthesis method can be applied to QAE when it is used to replace the Monte Carlo integration.

## ACKNOWLEDGMENT

This work was supported by Korea National Research Foundation (NRF) grant No. NRF-2020M3E4A1080031: Quantum circuit optimization for efficient quantum computing, NRF grant No. NRF-2020M3H3A1105796, ICT R&D program of MSIT/IITP 2021-0-01810, AFOSR grant FA2386-21-1-0089 and Amazon Web Services.

## APPENDIX A: Quantum phase estimation

QPE is an algorithm for finding an eigenvector of a unitary operator using QFT. Consider a unitary operator, eigenvalue, and eigenstate to be U, $e^{2\pi i\theta}$, and $|u\rangle$, respectively. Then, QPE can find $\theta$ if the state $|u\rangle$ is prepared and the controlled-U operators are implemented. The canonical QPE is implemented according to the following process. First, the state $|0\rangle^{\otimes t}|u\rangle$ is prepared, where t is a positive integer related to the precision of QPE. Second, apply t Hadamard (H) gates to $|0\rangle^{\otimes t}$. Third, apply the controlled-$U^j$ operator to the total state, where the controlled-$U^j$ operator transform $|j\rangle|u\rangle$ to $|j\rangle U^j|u\rangle$, where $|j\rangle$ is a computational basis state. Finally, implement the inverse Fourier transform on the first register and measure it. The measurement result gives a number that is an approximation of $2^t\theta$, which is accurate to $(t - \log_2(2 + 1/2\varepsilon))$ bits with a success probability of at least $(1 - \varepsilon)$ [11].

**APPENDIX B: Quantum amplitude estimation**

QAE is a frequently used sub-routine of quantum algorithms. An important feature of QAE is that it gives a quadratic speed-up compared to the classical Monte Carlo simulation [18].

QAE is an algorithm for finding the amplitude of a state $|\psi_1\rangle|1\rangle$ in the state $A|0\rangle^{\otimes(n+1)} = \sqrt{1-a}|\psi_0\rangle|0\rangle + \sqrt{a}|\psi_1\rangle|1\rangle$, where $A$ is a unitary operator. The canonical QAE is the QPE of the Grover operator $Q$ [Eq. (B1)]. Thus, the measurement result rightly converges to $O(1/M)$ with a probability of at least $8/\pi^2$, where $M$ is the number of qubits representing the measurement result [3].

$$Q(A, \chi) \equiv -AS_0 A^{-1} S_\chi \tag{B1}$$

$$S_0 \equiv I - 2|0\rangle\langle 0|^{\otimes(n+1)} \tag{B2}$$

$$S_\chi \equiv I - 2|\psi_1\rangle\langle\psi_1| \otimes |1\rangle\langle 1| \tag{B3}$$

Recently, QAEs that do not require the QPE have been proposed [22, 23]. They reduce the algorithmic costs compared with the canonical QAE because they do not use additional qubits, controlled operations, and the inverse QFT. However, the QAE without QPE, similar to the canonical QAE, uses the quantum amplitude amplification by the repetitive execution of the Grover operator $Q$. Therefore, reducing the implementation cost of the Grover operator $Q$ is considered a key to efficiently implementing QAE.

One of the most frequently appearing subcircuits in the circuit design of a Grover operator $Q$ is the circuit of serial controlled-$R_y$ gates sharing the target qubit. This is because it can express the basic approximation form of operator $A$ above when QAE is used to implement integration numerically [19]. The process is as follows. The goal of using QAE to implement integration numerically is to find $\sum f(x)$. To implement QAE, $A$ and $\theta(x)$ are defined as an equation below [Eq. (B4) and (B5)].

$$A|0\rangle_{n+1} = \sum \sqrt{(1-f(x))}|x\rangle_n|0\rangle + \sum \sqrt{f(x)}|x\rangle_n|1\rangle \qquad (B4)$$

$$\sqrt{f(x)} \equiv \sin(\theta(x)/2), x = \sum_{k=0}^{n} x_k 2^k, x_i = 0 \text{ or } 1 \qquad (B5)$$

Then, $\theta(x)$ can be written as $\sum_{j=0}^{n} a_j x^j = a_0 + x_0\theta_0 + x_1\theta_1 + \cdots + x_0 x_1 \theta_{01} + \cdots$, where each $\theta_k$ is a linear combination of $a_j$'s. Therefore, operator $A$ can be approximated to the required precision using H gates and multi-qubit controlled-$R_y$ gates sharing the target qubit. The basic approximation is the case n = 1, which can be implemented using a $R_y$ gate and controlled-$R_y$ gates sharing the target qubit. This basic approximation is practically useful for solving practical financial problems like risk analysis [19] or option pricing [21].

**REFERENCES**

[1] P. W. Shor, in Proceedings 35th annual symposium on foundations of computer science (IEEE 1994), p. 124.

[2] L. K. Grover, Quantum mechanics helps in searching for a needle in a haystack, Phys. Rev. Lett. 79, 325 (1997).

[3] G. Brassard, P. Hoyer, M. Mosca, and A. Tapp, Quantum amplitude amplification and estimation, Contemp. Math. 305, 53 (2002).

[4] A. W. Harrow, A. Hassidim, and S. Lloyd, Quantum algorithm for linear systems of equations, Phys. Rev. Lett. 103, 150502 (2009).

[5] I. M. Georgescu, S. Ashhab, and F. Nori, Quantum simulation, Rev. Mod. Phys. 86, 153 (2014).

[6] J.-H. Bae, P. M. Alsing, D. Ahn, and W. A. Miller, Quantum circuit optimization using quantum Karnaugh map, Sci. Rep. 10, 1 (2020).

[7] J. Preskill, Quantum computing in the NISQ era and beyond, Quantum 2, 79 (2018).

[8] A. Y. Kitaev, Quantum computations: algorithms and error correction, Russ. Math. Surv. 52, 1191 (1997).

[9] T. G. Draper, Addition on a quantum computer, arXiv:quant-ph/0008033 (2000).

[10] A. Y. Kitaev, Quantum measurements and the Abelian stabilizer problem, arXiv:quant-ph/9511026 (1995).

[11] M. A. Nielsen and I. L. Chuang, *Quantum Computation and Quantum Information*, 10th ed. (Cambridge University Press, Cambridge, 2010).

[12] A. Barenco, C. H. Bennett, R. Cleve, D. P. DiVincenzo, N. Margolus, P. Shor, T. Sleator, J. A. Smolin, and H. Weinfurter, Elementary gates for quantum computation, Phys. Rev. A 52, 3457 (1995).

[13] M. Saeedi, R. Wille, and R. Drechsler, Synthesis of quantum circuits for linear nearest neighbor architectures, Quantum Inf. Process. 10, 355 (2011).

[14] R. Wille, A. Lye, and R. Drechsler, Exact reordering of circuit lines for nearest neighbor quantum architectures, IEEE Trans. Comput. Aided Design Integr. Circuits Syst. 33, 1818 (2014).

[15] A. Kole, K. Datta, and I. Sengupta, A new heuristic for N-dimensional nearest neighbor realization of a quantum circuit, IEEE Trans. Comput. Aided Design Integr. Circuits Syst. 37, 182 (2017).

[16] A. Bhattacharjee, C. Bandyopadhyay, R. Wille, R. Drechsler, and H. Rahaman, in 2019 32nd International Conference on VLSI Design and 2019 18th International Conference on Embedded Systems (VLSID) (IEEE 2019), p. 203.

[17] A. G. Fowler, S. J. Devitt, and L. C. L. Hollenberg, Implementation of Shor's algorithm on a linear nearest neighbour qubit array, Quantum Inform. Comput. 4, 237 (2004).

[18] A. Montanaro, Quantum speedup of Monte Carlo methods, Proc. R. Soc. A-Math. Phys. Eng. Sci. 471, 20150301 (2015).

[19] S. Woerner and D. J. Egger, Quantum risk analysis, npj Quantum Inform. 5, 1 (2019).

[20] B. Park and D. Ahn, Halving the cost of quantum Fourier transform, arXiv:2203.07739 (2022).

[21] N. Stamatopoulos, D. J. Egger, Y. Sun, C. Zoufal, R. Iten, N. Shen, and S. Woerner, Option pricing using quantum computers, Quantum 4, 291 (2020).

[22] Y. Suzuki, S. Uno, R. Raymond, T. Tanaka, T. Onodera, and N. Yamamoto, Amplitude estimation without phase estimation, Quantum Inf. Process. 19, 1 (2020).

[23] D. Grinko, J. Gacon, C. Zoufal, and S. Woerner, Iterative quantum amplitude estimation, npj Quantum Inform. 7, 1 (2021).

**FIGURE AND TABLE CAPTIONS**

FIG. 1. 5-qubit quantum Fourier transform (QFT) circuits in the process. (a) The standard 5-qubit QFT circuit [11]. (b) The decomposed 5-qubit QFT circuit using the method in Ref. [20].

FIG. 2. A circuit identity. This circuit identity holds for all n qubits not only for 5-qubits, and is used to make the decomposed quantum Fourier transform circuit in Fig. 1(b) into a one-dimensional nearest-neighbor circuit in Fig. 3.

FIG. 3. 5-qubit nearest-neighbor quantum Fourier transform (QFT) circuit. Using the same form, a one-dimensional nearest-neighbor n-qubit QFT circuit can be constructed given $n^2 + n - 4$ CNOT gates.

Table I. The number of CNOT gates of the quantum Fourier transform (QFT) circuit with a one-dimensional nearest-neighbor structure. The first column represents the number of qubits of the QFT circuit, the second column represents our result, the third to the fifth columns represent the result of previous studies [15-17], and the sixth column represents the improvement rate of our circuit compared with the best-known result.

FIG. 4. Connectivity between qubits. (a) Partial circuit diagram of Rigetti-Aspen-11, showing connectivity between its qubits. The qubits labeled 10, 11, 26, and 27 are used to implement quantum phase estimation (QPE). (b) Circuit diagram of ibm_nairobi, showing connectivity between its qubits. The qubits labeled 1, 3, 5, and 4 are used to implement QPE.

FIG. 5. The result and comparison of the implementation of 3-qubit quantum phase estimation (QPE) on the Rigetti-Aspen-11 and ibm_nairobi. The blue and yellow columns are the results of Rigetti-Aspen-11 and ibm_nairobi implementations, respectively. The x-axis except the last one displays the $\theta$ that is the goal for QPE to find. QPE was implemented 1000 times for each $\theta$. The y-axis displays how many times the right answer $\theta$ was found. The last columns show the averages of how many times the right answers are obtained.

FIG. 1.

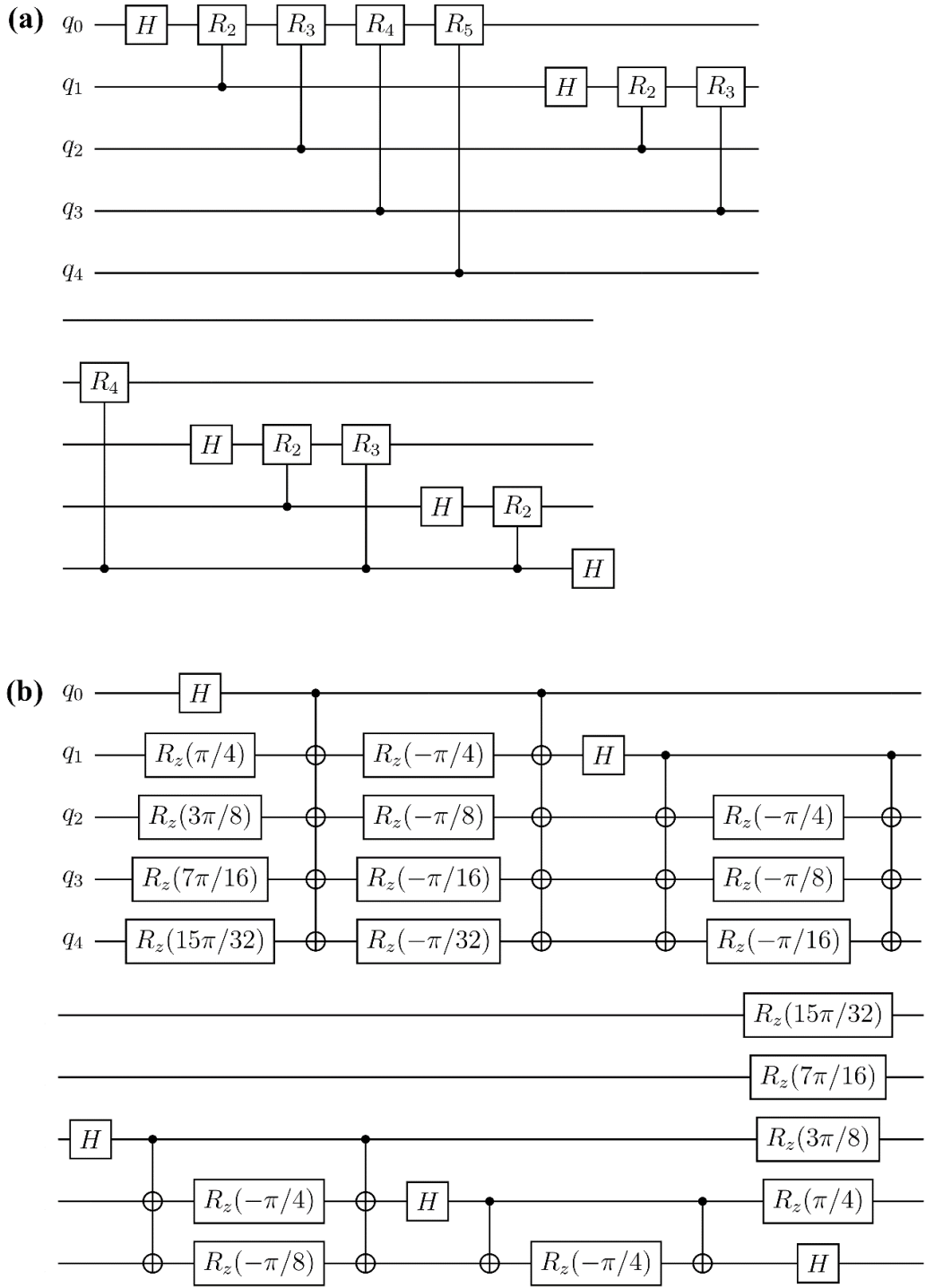

FIG. 2.

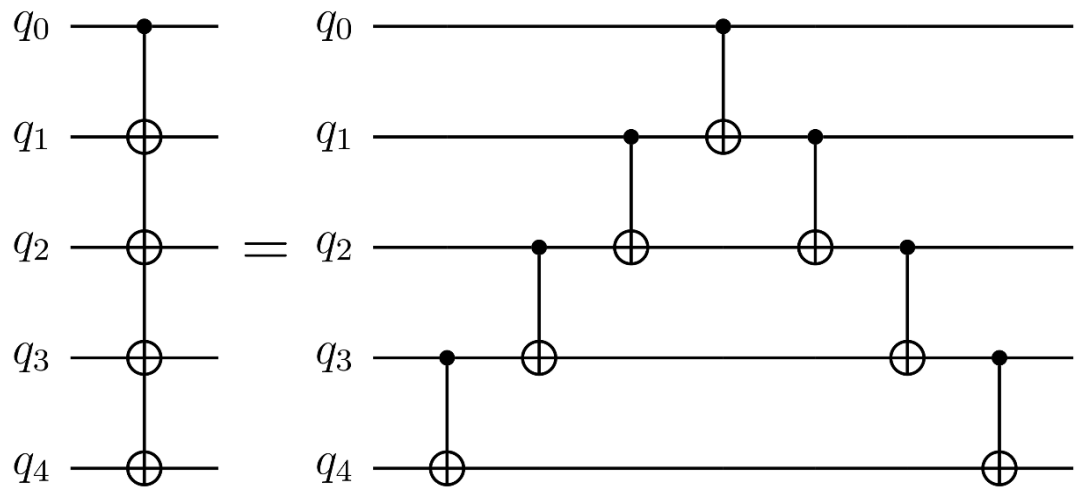

FIG. 3.

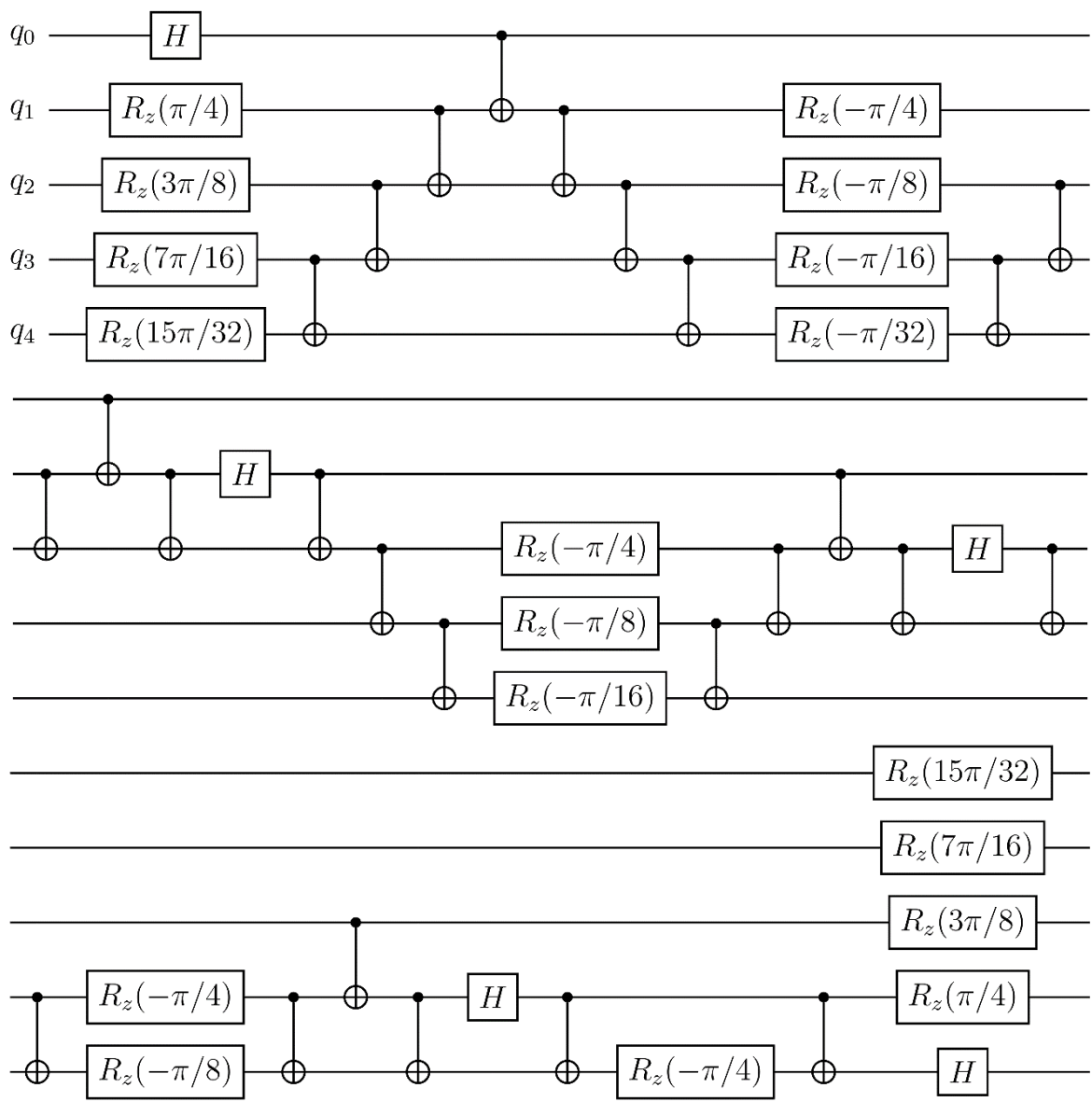

Table I.

|   | Ours | Ref. [17] | Ref. [15] | Ref. [16] | Improvement |
|---|---|---|---|---|---|
| n | $n^2+n-4$ | $(5/2)(n^2-1)$ | - | - | ~ 60% |
| 5 | 26 | 50 | 31 | - | ~ 16.13% |
| 6 | 38 | 75 | 48 | - | ~ 20.83% |
| 7 | 52 | 105 | 105 | 75 | ~ 30.67% |
| 8 | 68 | 140 | 124 | 121 | ~ 43.80% |
| 9 | 86 | 180 | 192 | 165 | ~ 47.88% |
| 10 | 106 | 225 | 240 | 225 | ~ 52.89% |

FIG. 4.

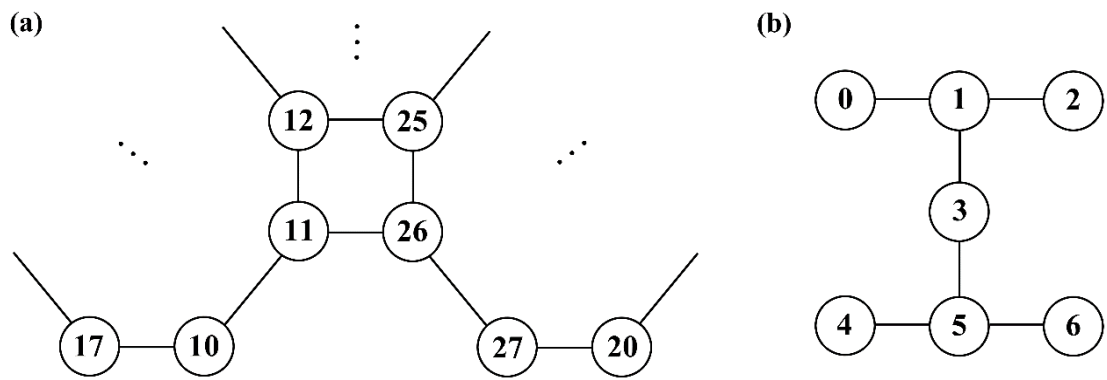

FIG. 5.

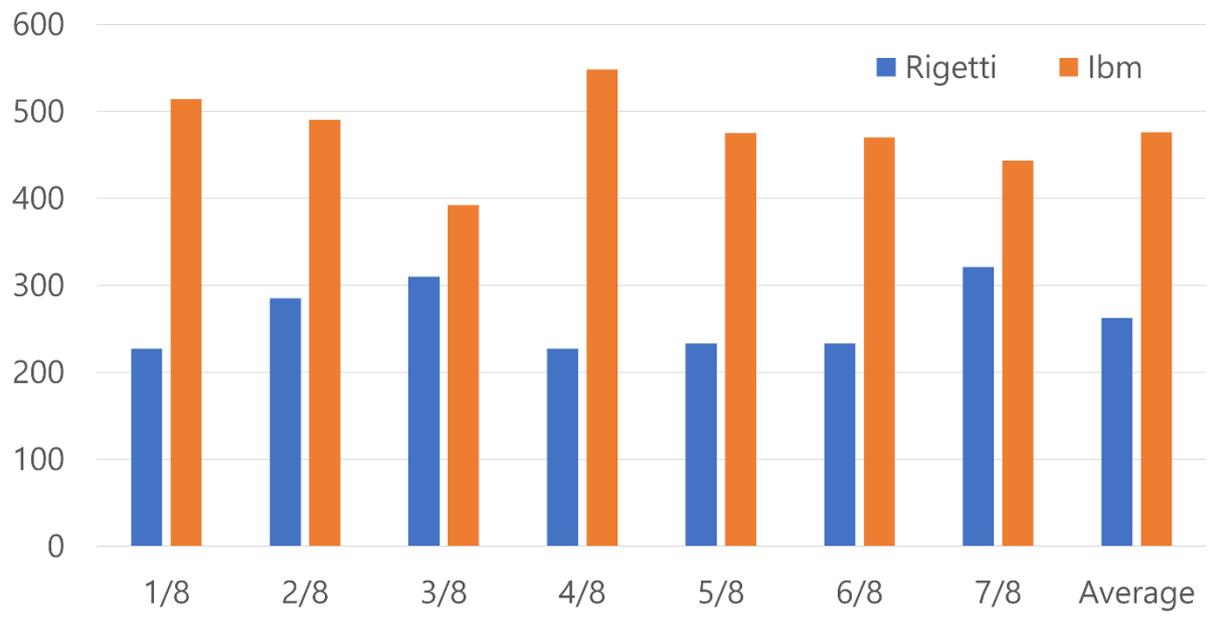